\documentclass[aps,prl,showpacs,twocolumn,lineno,groupedaddress,nofootinbib]{revtex4}  
\usepackage{graphicx}  
\usepackage{dcolumn}   
\usepackage{bm}        
\usepackage{amssymb}   

\begin{document}


\hspace{5.2in} \mbox{FERMILAB-Pub-04/286-E}

\title{Measurement of the $\Lambda^0_b$ lifetime in the decay $\Lambda^0_b \to J/\psi
\Lambda^0$ with the D\O\ Detector}
%
\author{                                                                      
V.M.~Abazov,$^{33}$                                                           
B.~Abbott,$^{70}$                                                             
M.~Abolins,$^{61}$                                                            
B.S.~Acharya,$^{27}$                                                          
M.~Adams,$^{48}$                                                              
T.~Adams,$^{46}$                                                              
M.~Agelou,$^{17}$                                                             
J.-L.~Agram,$^{18}$                                                           
S.H.~Ahn,$^{29}$                                                              
M.~Ahsan,$^{55}$
% ...
\\                                                                            
\vskip 0.30cm                                                                 
\centerline{(D\O\ Collaboration)}                                             
\vskip 0.30cm                                                                 
}                                                                             
\address{                                                                     
\centerline{$^{1}$Universidad de Buenos Aires, Buenos Aires, Argentina}       
\centerline{$^{2}$LAFEX, Centro Brasileiro de Pesquisas F{\'\i}sicas,         
                  Rio de Janeiro, Brazil}                                     
\centerline{$^{3}$Universidade do Estado do Rio de Janeiro,                   
                  Rio de Janeiro, Brazil}                                     
\centerline{$^{4}$Instituto de F\'{\i}sica Te\'orica, Universidade            
                  Estadual Paulista, S\~ao Paulo, Brazil}                     
\centerline{$^{5}$Simon Fraser University, Burnaby, Canada, University of     
                  Alberta, Edmonton, Canada,}                                 
\centerline{McGill University, Montreal, Canada and York University,          
                  Toronto, Canada}                                            
\centerline{$^{6}$Institute of High Energy Physics, Beijing,                  
                  People's Republic of China}                                 
\centerline{$^{7}$Universidad de los Andes, Bogot\'{a}, Colombia}             
\centerline{$^{8}$Charles University, Center for Particle Physics,            
                  Prague, Czech Republic}                                     
\centerline{$^{9}$Czech Technical University, Prague, Czech Republic}         
\centerline{$^{10}$Institute of Physics, Academy of Sciences, Center          
                  for Particle Physics, Prague, Czech Republic}               
\centerline{$^{11}$Universidad San Francisco de Quito, Quito, Ecuador}        
\centerline{$^{12}$Laboratoire de Physique Corpusculaire, IN2P3-CNRS,         
                 Universit\'e Blaise Pascal, Clermont-Ferrand, France}        
\centerline{$^{13}$Laboratoire de Physique Subatomique et de Cosmologie,      
                  IN2P3-CNRS, Universite de Grenoble 1, Grenoble, France}     
\centerline{$^{14}$CPPM, IN2P3-CNRS, Universit\'e de la M\'editerran\'ee,     
                  Marseille, France}                                          
\centerline{$^{15}$Laboratoire de l'Acc\'el\'erateur Lin\'eaire,              
                  IN2P3-CNRS, Orsay, France}                                  
\centerline{$^{16}$LPNHE, Universit\'es Paris VI and VII, IN2P3-CNRS,         
                  Paris, France}                                              
\centerline{$^{17}$DAPNIA/Service de Physique des Particules, CEA, Saclay,    
                  France}                                                     
\centerline{$^{18}$IReS, IN2P3-CNRS, Universit\'e Louis Pasteur, Strasbourg,  
                  France and Universit\'e de Haute Alsace, Mulhouse, France}  
\centerline{$^{19}$Institut de Physique Nucl\'eaire de Lyon, IN2P3-CNRS,      
                   Universit\'e Claude Bernard, Villeurbanne, France}         
\centerline{$^{20}$RWTH Aachen, III. Physikalisches Institut A,               
                   Aachen, Germany}                                           
\centerline{$^{21}$Universit{\"a}t Bonn, Physikalisches Institut,             
                  Bonn, Germany}                                              
\centerline{$^{22}$Universit{\"a}t Freiburg, Physikalisches Institut,         
                  Freiburg, Germany}                                          
\centerline{$^{23}$Universit{\"a}t Mainz, Institut f{\"u}r Physik,            
                  Mainz, Germany}                                             
\centerline{$^{24}$Ludwig-Maximilians-Universit{\"a}t M{\"u}nchen,            
                   M{\"u}nchen, Germany}                                      
\centerline{$^{25}$Fachbereich Physik, University of Wuppertal,               
                   Wuppertal, Germany}                                        
\centerline{$^{26}$Panjab University, Chandigarh, India}                      
\centerline{$^{27}$Tata Institute of Fundamental Research, Mumbai, India}     
\centerline{$^{28}$University College Dublin, Dublin, Ireland}                
\centerline{$^{29}$Korea Detector Laboratory, Korea University,               
                   Seoul, Korea}                                              
\centerline{$^{30}$CINVESTAV, Mexico City, Mexico}                            
\centerline{$^{31}$FOM-Institute NIKHEF and University of                     
                  Amsterdam/NIKHEF, Amsterdam, The Netherlands}               
\centerline{$^{32}$University of Nijmegen/NIKHEF, Nijmegen, The               
                  Netherlands}                                                
\centerline{$^{33}$Joint Institute for Nuclear Research, Dubna, Russia}       
\centerline{$^{34}$Institute for Theoretical and Experimental Physics,        
                  Moscow, Russia}                                             
\centerline{$^{35}$Moscow State University, Moscow, Russia}                   
\centerline{$^{36}$Institute for High Energy Physics, Protvino, Russia}       
\centerline{$^{37}$Petersburg Nuclear Physics Institute,                      
                   St. Petersburg, Russia}                                    
\centerline{$^{38}$Lund University, Lund, Sweden, Royal Institute of          
                   Technology and Stockholm University, Stockholm,            
                   Sweden and}                                                
\centerline{Uppsala University, Uppsala, Sweden}                              
\centerline{$^{39}$Lancaster University, Lancaster, United Kingdom}           
\centerline{$^{40}$Imperial College, London, United Kingdom}                  
\centerline{$^{41}$University of Manchester, Manchester, United Kingdom}      
\centerline{$^{42}$University of Arizona, Tucson, Arizona 85721, USA}         
\centerline{$^{43}$Lawrence Berkeley National Laboratory and University of    
                  California, Berkeley, California 94720, USA}                
\centerline{$^{44}$California State University, Fresno, California 93740, USA}
\centerline{$^{45}$University of California, Riverside, California 92521, USA}
\centerline{$^{46}$Florida State University, Tallahassee, Florida 32306, USA} 
\centerline{$^{47}$Fermi National Accelerator Laboratory, Batavia,            
                   Illinois 60510, USA}                                       
\centerline{$^{48}$University of Illinois at Chicago, Chicago,                
                   Illinois 60607, USA}                                       
\centerline{$^{49}$Northern Illinois University, DeKalb, Illinois 60115, USA} 
\centerline{$^{50}$Northwestern University, Evanston, Illinois 60208, USA}    
\centerline{$^{51}$Indiana University, Bloomington, Indiana 47405, USA}       
\centerline{$^{52}$University of Notre Dame, Notre Dame, Indiana 46556, USA}  
\centerline{$^{53}$Iowa State University, Ames, Iowa 50011, USA}              
\centerline{$^{54}$University of Kansas, Lawrence, Kansas 66045, USA}         
\centerline{$^{55}$Kansas State University, Manhattan, Kansas 66506, USA}     
\centerline{$^{56}$Louisiana Tech University, Ruston, Louisiana 71272, USA}   
\centerline{$^{57}$University of Maryland, College Park, Maryland 20742, USA} 
\centerline{$^{58}$Boston University, Boston, Massachusetts 02215, USA}       
\centerline{$^{59}$Northeastern University, Boston, Massachusetts 02115, USA} 
\centerline{$^{60}$University of Michigan, Ann Arbor, Michigan 48109, USA}    
\centerline{$^{61}$Michigan State University, East Lansing, Michigan 48824,   
                   USA}                                                       
\centerline{$^{62}$University of Mississippi, University, Mississippi 38677,  
                   USA}                                                       
\centerline{$^{63}$University of Nebraska, Lincoln, Nebraska 68588, USA}      
\centerline{$^{64}$Princeton University, Princeton, New Jersey 08544, USA}    
\centerline{$^{65}$Columbia University, New York, New York 10027, USA}        
\centerline{$^{66}$University of Rochester, Rochester, New York 14627, USA}   
\centerline{$^{67}$State University of New York, Stony Brook,                 
                   New York 11794, USA}                                       
\centerline{$^{68}$Brookhaven National Laboratory, Upton, New York 11973, USA}
\centerline{$^{69}$Langston University, Langston, Oklahoma 73050, USA}        
\centerline{$^{70}$University of Oklahoma, Norman, Oklahoma 73019, USA}       
\centerline{$^{71}$Brown University, Providence, Rhode Island 02912, USA}     
\centerline{$^{72}$University of Texas, Arlington, Texas 76019, USA}          
\centerline{$^{73}$Southern Methodist University, Dallas, Texas 75275, USA}   
\centerline{$^{74}$Rice University, Houston, Texas 77005, USA}                
\centerline{$^{75}$University of Virginia, Charlottesville, Virginia 22901,   
                   USA}                                                       
\centerline{$^{76}$University of Washington, Seattle, Washington 98195, USA}  
}                                                                             
\date{\today}

\begin{abstract}

We present measurements of the $\Lambda^0_b$ lifetime in the exclusive decay
channel $\Lambda^0_{b}\rightarrow J/\psi \Lambda^0$, 
with $J/\psi \to \mu^+ \mu^-$ and $\Lambda^0 \to p \pi^-$, the $B^0$ lifetime 
in the decay $B^0 \rightarrow J/\psi K^0_S$ with $J/\psi \to \mu^+ \mu^-$ 
and $K^0_S \to \pi^+ \pi^-$, and the ratio of these lifetimes. 
The analysis is based on approximately 250 pb$^{-1}$ of
data recorded with the D\O\ detector in $p \bar{p}$ collisions at
$\sqrt{s}$=1.96~TeV.
The $\Lambda^0_b$ lifetime 
is determined to be $\tau(\Lambda^0_b) = 1.22^{+0.22}_{-0.18}\mbox{(stat)}\pm 0.04 \mbox{(syst)}$ ps, 
the $B^0$ lifetime $\tau(B^0) = 1.40^{+0.11}_{-0.10}\mbox{(stat)}\pm 0.03 \mbox{(syst)}$ ps, 
and the ratio $\tau(\Lambda^0_b)/\tau(B^0) = 0.87^{+0.17}_{-0.14}\mbox{(stat)}\pm 0.03\mbox{(syst)}$.  
In contrast with previous measurements using semileptonic decays, this is the first determination 
of the $\Lambda^0_b$ lifetime based on a fully reconstructed decay channel.

\end{abstract}

\pacs{14.20.Mr, 14.40.Nd, 13.30.Eg, 13.25.Hw}
\maketitle

\vskip.5cm 

\newpage

Calculations based on a simple quark-spectator model~\cite{spect} predict
that the lifetimes of all $b$ hadrons are equal.  When non-spectator effects
are taken into account, they give rise to a lifetime hierarchy of $\tau(B^+)
\ge \tau(B^0) \approx \tau(B^0_s) > \tau(\Lambda^0_b) \gg
\tau(B^+_c)$~\cite{Bdiff}.  Measurements of $b$-hadron lifetimes therefore
provide means to determine the importance of non-spectator contributions in
$b$-hadron decays.  For comparison with theory, measurements of lifetime
ratios are preferred over individual lifetimes.
The ratio $\tau(\Lambda^0_b)/\tau(B^0)$ has been the source of theoretical
study since early calculations~\cite{Lblf1} predicted a value greater than
0.90, almost two sigma away from the measurement average~\cite{PDG2004},
$0.800\pm0.053$. Recent calculations~\cite{recLb} of this ratio that include
higher order effects have reduced this difference. A current
compilation~\cite{PDG2004} of lifetime ratio data for $b$ hadrons
is consistent with early calculations~\cite{teopred} that include
non-spectator effects.
Previous measurements of $\tau(\Lambda^0_b)$ used
semileptonic decay channels that suffer from uncertainties arising from
undetected neutrinos. 
A measurement of the lifetime using fully reconstructed $\Lambda^0_b$ decays
is free from the neutrino ambiguities.
The Tevatron Collider at Fermilab is the only operating accelerator 
where $\Lambda^0_b$ baryons are being produced and studied.

In this Letter we report a measurement of the $\Lambda^0_{b}$ lifetime in the
decay channel $\Lambda^0_{b}\rightarrow J/\psi \Lambda^0$, and its ratio to
the $B^0$ lifetime from the $B^0\rightarrow J/\psi K^{0}_{S}$ decay channel. 
This $B^0$ decay channel is chosen because of its similar topology to
the $\Lambda^0_b$ decay.  The $J/\psi$ is
reconstructed in the $\mu^+ \mu^-$ decay mode, the $\Lambda^0$ in
$p\pi^{-}$, and the $K^0_S$ in $\pi^+ \pi^-$; throughout this
Letter the appearence of a specific charge state will also imply its charge
conjugate. The data used in this analysis were collected during 2002--2004
with the D\O\ detector in Run II of the Tevatron Collider at a
center-of-mass energy of 1.96 TeV, and correspond to an integrated
luminosity of approximately 250~pb$^{-1}$.


The components of the D\O\ detector~\cite{run2det} most relevant for this measurement
are the charged-particle tracking systems and the muon detector.
The D\O\
tracker consists of a silicon microstrip tracker (SMT) and a central fiber
tracker (CFT) that are surrounded by a superconducting solenoid magnet that
produces a 2~T central magnetic field. The SMT has approximately 800,000
individual strips, with a typical pitch of $50-80$ $\mu$m, and a design
optimized for tracking and vertexing capability for $|\eta|<3$
($\eta = -\ln[\tan(\theta/2)]$ and $\theta$ is the polar angle).  
The system has a six-barrel longitudinal structure interspersed with sixteen
disks. Each barrel consists of four layers arranged axially around the beam
and the disks are placed perpendicular to the beam.
The CFT has eight thin coaxial barrels, each
supporting two doublets of overlapping scintillating fibers of 0.835~mm
diameter, one doublet being parallel to the collision axis, and the other
alternating by $\pm 3^{\circ}$ relative to the axis.  
For charged particles, the resolution for the distance of closest approach
as provided by the tracking system is approximately 50~$\mu$m for tracks
with $p_T \approx 1$~GeV/$c$, and improves asymptotically to 15~$\mu$m for
tracks with $p_T \geq 10$~GeV/$c$, $p_T$ is the component of the momentum
perpendicular to the beam pipe. Preshower detectors and electromagnetic
and hadronic calorimeters surround the tracker.
A muon system is located beyond the calorimeter, and consists of 
multilayer drift chambers
and scintillation trigger counters preceding 1.8~T
toroidal magnets, followed by two similar layers beyond the toroids.
Muon identification for $|\eta|<1$ relies on 10~cm wide drift
tubes, while 1~cm mini-drift tubes are used for $1<|\eta|<2$.

Primary vertex (PV) candidates are determined for each event by minimizing a
$\chi^2$ function that depends on all the tracks in the event and a term
that represents the beam spot constraint. The beam spot is the run-by-run
average beam position, where a run typically lasts several hours.  The beam
spot is stable during the periods of time when the proton and antiproton
beams are kept colliding continuously and can be used as a constraint for
the primary vertex fit. The initial primary vertex candidate and its
$\chi^2$ are obtained using all tracks. Next, each track used in the
$\chi^2$ calculation is removed temporarily and the $\chi^2$ is calculated
again; if the $\chi^2$ decreases by 9 or more, this track is discarded from
the PV fit. This procedure is repeated until no more tracks can be
discarded.  Additional primary vertices are obtained by applying the same
algorithm to the discarded tracks until no more vertices are found.

We base our data selection on defined objects such as charged tracks and
muons. Although we do not require any specific trigger to select our sample,
most of the events selected fire dimuon or single muon triggers. Preliminary
selection of dimuon events requires the presence of at least two muons of
opposite charge reconstructed in the tracker and the muon system.  We
require that at least one of the muon candidates consists of a central
track, defined by hits in the SMT and CFT, matched with muon track segments
in all three layers of the muon system. For the second muon, we require a
central track matched with hits in at least the innermost layer of the muon
system. The sample of $J/\psi \to \mu^+ \mu^-$ candidates consists of events
with at least two muons, with trajectories constrained in a fit to a common
vertex. The fit must have a $\chi^2$ probability greater than 1\%, and the
invariant mass of the dimuons must be in the range $2.80 < M_{\mu \mu} < 3.35$~GeV/$c^2$. 
To reconstruct $\Lambda^0_b$ and $B^0$ candidates, the
$J/\psi$ events are examined for $\Lambda^0$ and $K^0_S$ candidates. 
The $\Lambda^0 \to p \pi^-$ candidates are required to have two tracks of
opposite charge which must originate from a common vertex with a $\chi^2$
probability greater than 1\%. A candidate is selected if the mass of the
proton-pion system after the vertex-constrained fit falls in the $1.100 <
M_{p\pi} < 1.128$ GeV/$c^2$ window.  The proton mass is assigned to the
track of higher momentum, and the $p_T$ of the $\Lambda^0$ is required to be
greater than 2.4 GeV/$c$. The $K^0_S \to \pi^+ \pi^-$ selection follows the
same criteria, except that the mass window is $0.460 < M_{\pi\pi} <
0.525$~GeV/$c^2$, and $p_T>$1.8~GeV/$c$.

We reconstruct the $\Lambda^0_b$ and $B^0$ by performing a constrained fit
to a common vertex for either the $\Lambda^0$ or $K^0_S$ and the two
muon tracks, with the latter constrained to the $J/\psi$ mass of 3.097
GeV/$c^2$~\cite{PDG2004}. Because of their long decay lengths, a significant 
fraction of $\Lambda^0$ and $K^0_S$ will decay outside the SMT.
Therefore, to maintain good efficiency, no SMT hits are required
on the tracks of the decay particles. To reconstruct the
$\Lambda^0_b$ ($B^0$), we first find the $\Lambda^0$ ($K^0_S$) decay
vertex, and then extrapolate the momentum vector of the ensuing
particle and form a vertex with it and the two muon tracks belonging to the
$J/\psi$.  The precision of the $\Lambda^0_b$ ($B^0$) vertex position is 
dominated by the two muon tracks from the $J/\psi$.  If more than one
candidate is found in the event, the candidate with the best $\chi^2$
probability is selected as the $\Lambda^0_b$ ($B^0$) candidate.


We determine the lifetime of a $\Lambda^0_b$ or $B^0$ by measuring the
distance traveled by each $b$-hadron candidate in a plane transverse to the
beam direction, and then applying a correction for the Lorentz boost.
We define the transverse decay length as 
$L_{xy} = \bm{L}_{xy} \cdot \bm{p}_T/p_T$ 
where $\bm{L}_{xy}$ is the vector that points
from the primary to the secondary vertex and $\bm{p}_T$ is the
transverse momentum vector of the $b$ hadron. The proper decay length
(PDL) for a $b$-hadron candidate is then given by:

\begin{equation}
\label{ctau}
{\rm PDL} =\frac{L_{xy}}{(\beta\gamma)_{T}^{B}} = L_{xy}\, \frac{c M_{B}}{p_{T}},
\end{equation}

\noindent where $(\beta\gamma)_{T}^{B}$, and $M_{B}$ are the transverse 
boost and the mass of the $b$ hadron, respectively.  In our
measurement, the value of $M_{B}$ in Eq. \ref{ctau} is set to the PDG mass
value of $\Lambda^0_{b}$ or $B^0$~\cite{PDG2004}. In our final selection of
$\Lambda^0_b$ and $B^0$ candidates, we require an error of less than 
100 $\mu$m on the PDL and we also require a total momentum greater than 5~GeV$/c$.

\begin{figure}
\includegraphics[scale=.45]{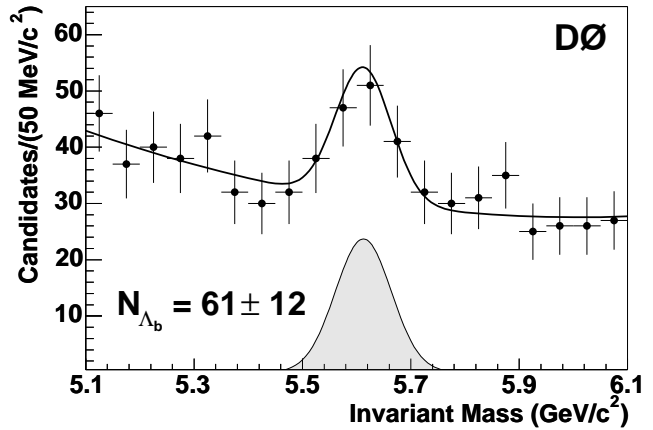}
\caption{\label{fig:lbm} Invariant mass distribution for $\Lambda^0_b$ candidate events.
The points represent the data, and the curve represents the result of the fit.
The mass distribution for the signal is shown in gray.}
\end{figure}
\begin{figure}
\includegraphics[scale=.45]{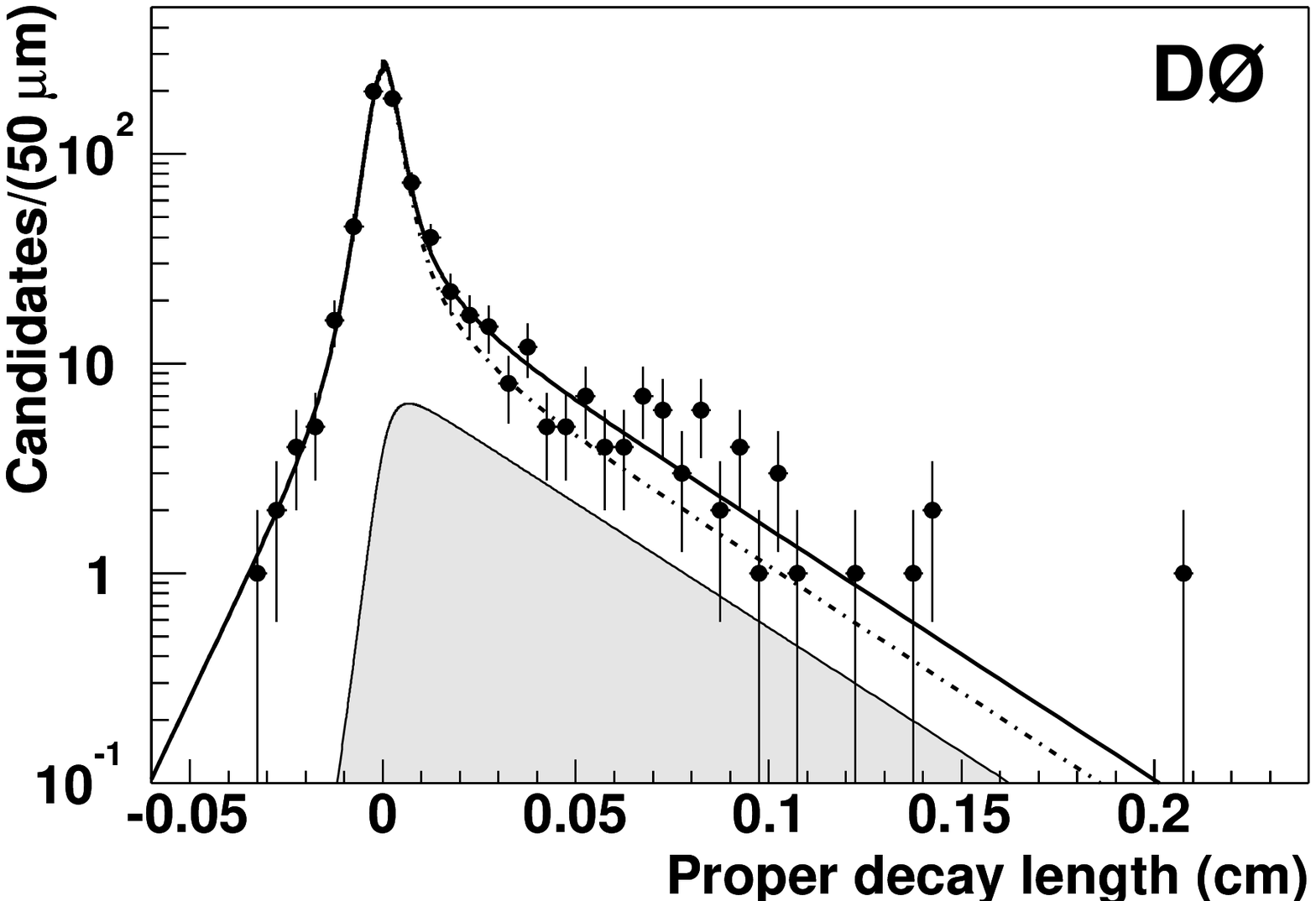}
\caption{\label{fig:lblt} Distribution of proper decay length for $\Lambda^0_b$
candidates. The points are the data, and the solid curve is the sum of the
contributions from signal (gray) and the background (dashed-dotted line).}
\end{figure}
\begin{figure}
\includegraphics[scale=.45]{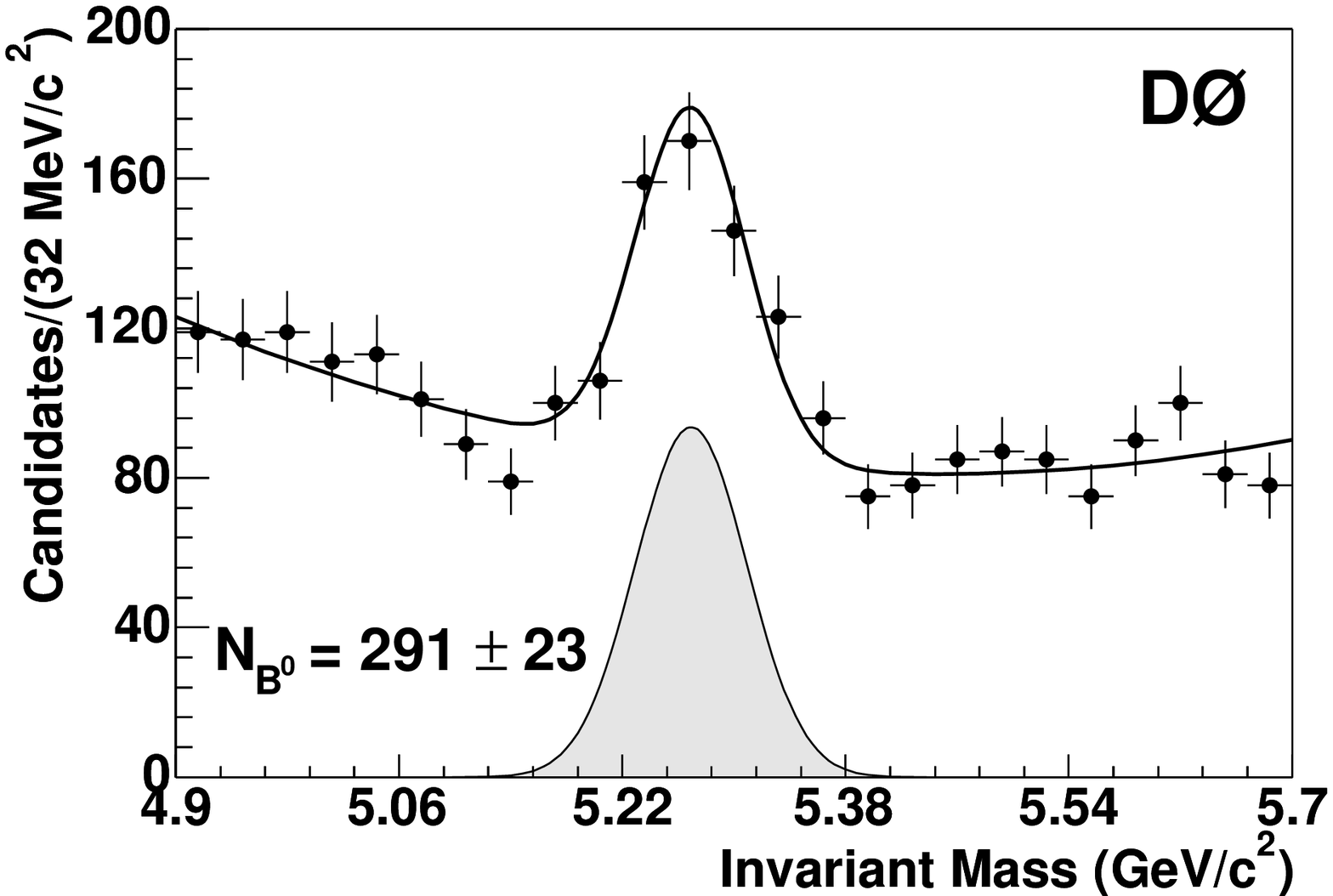}
\caption{\label{fig:b0m} Invariant mass distribution for $B^0$ candidate events.
The points represent the data, and the curve represents the result of the fit.
The mass distribution for the signal is shown in gray.}
\end{figure}
\begin{figure}
\includegraphics[scale=.45]{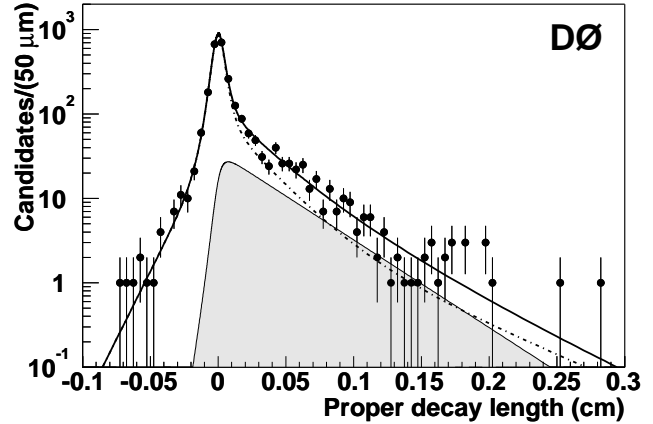}
\caption{\label{fig:b0lt} Distribution of proper decay length for $B^0$
candidates. The points are the data, and the solid curve is the sum of the
contributions from signal (gray) and the background (dashed-dotted line).}
\end{figure}

We perform an unbinned likelihood fit to measure the $\Lambda^0_b$ and $B^0$
lifetimes. The inputs for the fit are the mass, PDL and PDL error of the
candidates. Candidates with invariant masses in the range of 5.1 to 6.1
GeV/$c^2$ for the $\Lambda^0_b$ and 4.9 to 5.7 GeV/$c^2$ for the $B^0$ are
selected; these ranges include sideband regions that are used to model the PDL
distributions of backgrounds. The likelihood function, ${\cal L}$, is
defined by:

\begin{equation}
\begin{array}{lcl}
{\cal L} & = & \prod_{j=1}^{N} \left[ f_s S_M (M_j) S_{L} (\lambda_j, \sigma_j) \right.
\\
         & +  & \left. (1 - f_s) B_M (M_j) B_{L} (\lambda_j, \sigma_j)\right],
\end{array}
\end{equation}

\noindent where $N$ is the total number of selected events, $f_s$ is the fraction of 
signal events in the sample,  $S_M$ and
$B_M$ are the probability distribution functions used to model the mass
distributions for signal and background, respectively, and $S_{L}$ and $B_{L}$
model the distributions of proper decay lengths for signal and background.
The mass for signal is modeled by a Gaussian distribution and the
mass for background is described by a second-order polynomial.
The PDL distribution for signal is described by the convolution of an
exponential decay, whose decay constant is one of the parameters of the fit,
with a resolution function represented by a single Gaussian function:

\begin{equation}
G(\lambda_{j},\sigma_{j}) =
\frac{1}{\sqrt{2\pi}s\sigma_{j}} \exp \left(\frac{-\lambda_{j}^{2}}{2(s\sigma_j)^{2}}\right),
\label{res}
\end{equation}

\noindent where $\lambda_j$ and $\sigma_j$ represent the PDL and its error
respectively for a given event $j$,
and the $s$ parameter is introduced in the fit to account for a possible
misestimate of $\sigma_j$. The PDL distribution
for background is described by a sum of a resolution function representing
the zero-lifetime component, negative and positive exponential decay
functions modeling combinatorial background, and an exponential decay that
accounts for long-lived heavy flavor decays. 
We minimize $-2\ln{{\cal L}}$
to extract the parameters: 
$c\tau (\Lambda^0_b) = 366^{+65}_{-54}$~$\mu$m 
and $c\tau(B^0) = 419^{+32}_{-29}$~$\mu$m. 
From the fits, we get $s = 1.27\pm 0.10$ and
$s=1.39\pm 0.05$ for the $\Lambda^0_b$ and $B^0$ respectively; and the number of
signal events $61\pm12$ $\Lambda^0_b$ and $291\pm23$ $B^0$.
Figs~\ref{fig:lbm} and \ref{fig:lblt} (Figs~\ref{fig:b0m} and
\ref{fig:b0lt}) show the mass and proper decay length distributions for the
$\Lambda^0_b$ ($B^0$) candidates, respectively, with the results of the fits
superimposed.

Table~\ref{tab:table2} summarizes the systematic uncertainties on our
measurements. 
The contribution from the uncertainty on the detector alignment is estimated
by reconstructing the $B^0$ sample with the positions of the SMT sensors
shifted outwards radially by the alignment error in the radial position of
the sensors and then fitting for the lifetime.
We estimate the systematic uncertainty due to the resolution on the PDL by using 
two Gaussian functions for the resolution model. 
The contribution to the systematic uncertainty from the model describing
background PDLs is studied by varying the parametrizations of the
different components: (i) the exponential functions are replaced by
exponentials convoluted with the resolution function of Eq.~\ref{res}, (ii)
a uniform background is added to account for outlier events (this has only a
negligible effect), and (iii) the positive and negative short-lived lifetime
components are forced to be symmetric. 
To study the systematic uncertainty due to the model for the mass distributions,
we vary the shapes of the mass distributions for signal and background. For
the signal, we use two Gaussian functions instead of a single one, and for the
background distribution, a linear function instead of the nominal quadratic
form. 

The lifetime of the long-lived component of the background varies with mass.
This results in an uncertainty in the decay constant of the background under
the mass peaks. We obtain the systematic uncertainty due to this effect by
modelling the long-lived background with two exponentials instead of a
single exponential. The decay constant of one of the two exponentials is
determined from a fit in the low-mass sideband, and the other decay
constant is determined from the high-mass sideband. The low-mass sideband is
defined as the mass window 4.900-5.149~GeV/$c^2$ for $B^0$ and 5.100-5.456~GeV/$c^2$ for
$\Lambda^0_b$ and the high-mass sideband as 5.389-5.700~GeV/$c^2$ 
and 5.768-6.100~GeV/$c^2$ respectively.
We perform the fit incorporating the linear combination of exponentials with the
decay constants fixed to the values obtained in the low- and high-mass
sidebands fits and allowing the coefficients of the linear combination
to float. The systematic uncertainty quoted is the difference between the values we get
from this fit and the nominal.

We also study the contamination of the $\Lambda^0_b$ sample by $B^0$ events
that pass the $\Lambda^0_b$ selection. From Monte Carlo studies, we estimate
that 19 $B^0$ events are reconstructed as $\Lambda^0_b$ events. The
invariant masses of the $B^0$ events entering the $\Lambda^0_b$ sample are
distributed almost uniformly across the entire mass range, and do not peak
at the $\Lambda^0_b$ mass.  Their proper decay lengths therefore tend to be
incorporated in our model of the long-lived heavy-flavor component of the
background.  To estimate the systematic uncertainty due to this
contamination, we fit the mass and proper decay length distributions of the
misidentified events in the MC samples, add this contribution to the
likelihood with fixed parameters, and perform the fit again. The difference
between the two results is quoted as the systematic uncertainty due to the
contamination.

The fitting procedure is tested for the presence of biases by generating
1000 Monte Carlo experiments, each with the same statistics as our data
samples. For the generated events, the PDL errors are taken from data, and
the mass and PDL distributions are described by the probability distribution
functions used in data, with parameters obtained from the fit.  The fits
performed on these Monte Carlo experiments indicate that there is no bias
inherent in the procedure.

We also perform several cross-checks of the lifetime measurements. In
particular, a fit is done where the background is modeled using only
sideband regions, the $J/\psi$ vertex is used instead of the $b$-hadron
vertex, the mass windows are varied, the reconstructed $b$-hadron mass is
used instead of the Particle Data Group~\cite{PDG2004} value, and the sample
is split into different pseudorapidity regions or different regions of
azimuth.  All results obtained with these variations are consistent with our
central values.

\begin{table}
\caption{\label{tab:table2} Summary of systematic uncertainties in the
measurement of $c\tau$ for $\Lambda^0_b$  and $B^0$  
and their ratio. 
The total uncertainties are also given combining individual uncertainties in quadrature.}
\begin{ruledtabular}
\begin{tabular}{lccc}
Source & $\Lambda^0_{b}$ ($\mu$m) & $B^0$ ($\mu$m)  & Ratio \\
\hline
Alignment             & 5.4  & 5.4  &  0.002 \\
Model for PDL resolution  & 6.7  & 2.7  &  0.010 \\
Model for PDL background  & 2.7  & 3.1  &  0.005 \\
Model for signal mass & 0.2  & 0.0  &  0.000 \\
Model for background mass & 2.5  & 6.2  &  0.007 \\
Long-lived components & 1.5  & 0.1  &  0.003 \\
Contamination         & 8.8  & 0.8  &  0.023 \\
\hline
{\bf Total}           & {\bf 12.9} & {\bf 9.2}  &  {\bf 0.028} \\
\end{tabular}
\end{ruledtabular}
\end{table}


The results of our measurement of the $\Lambda^0_b$ and $B^0$ lifetimes are
summarized as:

\begin{eqnarray}
\tau(\Lambda^0_b) = 1.22^{+0.22}_{-0.18} \mbox{ (stat)} \pm 0.04 \mbox{ (syst) ps,}\\ \nonumber
\tau(B^0) = 1.40^{+0.11}_{-0.10} \mbox{ (stat)} \pm 0.03 \mbox{ (syst) ps}.
\end{eqnarray}

\noindent These can be combined to determine the ratio of lifetimes:

\begin{equation}
\frac{\tau(\Lambda^0_{b})}{\tau(B^0)} = 0.87^{+0.17}_{-0.14} \mbox{ (stat)}\pm 0.03 \mbox{ (syst)}, 
\end{equation}

\noindent where we determine the systematic uncertainty of the ratio by varying each
parameter in the two samples simultaneously and quoting the deviation in the
ratio as the systematic uncertainty due to that source.

In conclusion, we have measured the $\Lambda^0_b$ lifetime in the fully
reconstructed exclusive decay channel $J/\psi \Lambda^0$. This is the first
time that this lifetime has been measured in an exclusive
channel. The measurement is
consistent with the world average, $1.229\pm0.080$ ps~\cite{PDG2004}, and the $\Lambda^0_b$
to $B^0$ ratio of lifetimes is also consistent with 
theoretical predictions~\cite{teopred,Lblf1,recLb}.

%
We thank the staffs at Fermilab and collaborating institutions, 
and acknowledge support from the 
Department of Energy and National Science Foundation (USA),  
Commissariat  \` a l'Energie Atomique and 
CNRS/Institut National de Physique Nucl\'eaire et 
de Physique des Particules (France), 
Ministry of Education and Science, Agency for Atomic 
   Energy and RF President Grants Program (Russia),
CAPES, CNPq, FAPERJ, FAPESP and FUNDUNESP (Brazil),
Departments of Atomic Energy and Science and Technology (India),
Colciencias (Colombia),
CONACyT (Mexico),
KRF (Korea),
CONICET and UBACyT (Argentina),
The Foundation for Fundamental Research on Matter (The Netherlands),
PPARC (United Kingdom),
Ministry of Education (Czech Republic),
Natural Sciences and Engineering Research Council and 
WestGrid Project (Canada),
BMBF and DFG (Germany),
A.P.~Sloan Foundation,
Research Corporation,
Texas Advanced Research Program,
and the Alexander von Humboldt Foundation.
%

\end{document}